\documentclass[aps,prd,twocolumn,showpacs,showkeys,superscriptaddress,nofootinbib, 10pt]{revtex4-2}

\usepackage[utf8]{inputenc} % Input encoding
\usepackage{amsmath, amssymb} % Math symbols and environments
\usepackage{mathrsfs}
\usepackage{graphicx, subcaption} % Figures and subfigures
\usepackage{float} % Float placement
\usepackage{color, caption} % Caption customization
\usepackage{natbib} % Bibliography support
\usepackage{lineno}
\usepackage{hyperref, url}
\begin{document}

\title{Emergent gravity from Michel flow with position dependent adiabatic index}

\author{Apashanka Das}
\email{apashankadas@hri.res.in}
\affiliation{Harish-Chandra Research Institute,  HBNI, Chhatnag Road, Jhunsi, Prayagraj - 211 019, India}

\author{Souvik Ghose}
\email{dr.souvikghose@gmail.com}
\affiliation{HECRC, University of North Bengal, Siliguri - 734013, West Bengal, India}

\author{Tapas K. Das}
\email{tapas@hri.res.in}
\affiliation{Harish-Chandra Research Institute,  HBNI, Chhatnag Road, Jhunsi, Prayagraj - 211 019, India}

\date{\today}

\begin{abstract}\noindent
Spherically symmetric, general relativistic Bondi accretion is known as
the Michel flow. The stationary integral transonic solutions for the Michel flow has been constructed for multi-component accretion described by an equation of state where the adiabatic index varies with the
radial distance along which the streamlines are studied, and the
corresponding phase portrait spanned by such radial distance and the
flow Mach number has been obtained. Borrowing the techniques used in the
dynamical systems theory, the nature of the transonic points of the
aforementioned flow has been classified. The steady state flow has been
perturbed to study the stability of the stationary solutions, and it has
been found that such flows are stable under the (linear) radial
perturbation. As a consequence of the stability analysis,  the
corresponding acoustic space time embedded within the accreting matter
has been obtained, and the horizon of the metric of such sonic space
time has been identified by constructing the causal structure with the
help of the Carter-Penrose diagrams. In this way, the accreting black
hole systems in the general relativistic set up has been investigated
from various different perspectives - from its astrophysical aspects,
from the dynamical systems point of view, as well as within the realm of
the classical analogue gravity phenomena.  
\end{abstract}

\maketitle

% PACS numbers and keywords

\maketitle

% Main content

\section{Introduction}
\label{sec:intro}
Perturbation of transonic fluid may lead to the formation of acoustic
geometry embedded inside the fluid flow, where a black hole-like acoustic
metric describes the propagation of the acoustic perturbation, and an
acoustic horizon may form which mimics the black hole event horizon.
Generation of such a black hole-like spacetime by perturbing a
dissipationless, irrotational, barotropic classical or quantum fluid opens
up the possibility of studying various kinematic properties of the general
theory of relativity in the laboratory setup on a terrestrial scale, and
such study of gravity as an emergent phenomenon is known as analogue
gravity.
Unruh~\cite{Unruh1981} and Moncrief~\cite{Moncrief1980} were the pioneers
in introducing this important field of study; later,
Visser~\cite{Visser1998} and Bili'{c}~\cite{Bilic1999} introduced the
comprehensive formalism of emergent gravity phenomena (hereafter, emergent
gravity and analogue gravity will be used synonymously), and eventually a
series of classical and quantum black hole analogue models have been
developed in the literature; see,
e.g.,~\cite{BarceloLiberatiVisser2005,Almeida2023} for exhaustive
discussions on the history of the development of academic research in the
field of analogue gravity and for details about various analogue models
introduced in the literature.

Among all such analogue models, the study of emergent gravity phenomena in
accreting black hole systems deserves special attention.
An accreting astrophysical black hole is considered as a natural example of
a classical analogue gravity model, since black hole accretion is
necessarily transonic~\cite{Liang1980}, and perturbation of the accreting
fluid results in the generation of analogue spacetime embedded within the
transonic accretion.
An accreting black hole as an analogue model is unique in the sense that
only for this model are both the actual (gravitational) and the analogue
(acoustic) horizons present simultaneously; see,
e.g.,~\cite{Das2004,Das2005,AbrahamBilicDas2006,Maity2022} and references
therein for further details about such models. It is to be noted that although the velocity potential is usually perturbed
to construct the equation of motion for the corresponding massless scalar
field while generating the acoustic spacetime in standard analogue
models~\cite{Visser1998}, the mass accretion rate has been used to carry out
the linear perturbation scheme for developing the acoustic spacetime for
accreting black holes, since the mass accretion rate is a physically
meaningful and observationally measurable quantity.
Such a perturbation scheme does not produce any ambiguity since it has been
shown that the perturbation of the velocity potential as well as the mass
accretion rate yield the same acoustic metric~\cite{Shaikh2017}.

In one of the recent works on the analogue properties of accreting black
hole systems, the linear perturbation of general relativistic spherical
accretion of irrotational, barotropic, hydrodynamic matter onto a
Schwarzschild black hole has been shown to produce a black hole-like
acoustic metric as a consequence of the linear stability analysis of the
stationary integral accretion solutions.
Such works were performed for flow described by the polytropic equation of
state, where the adiabatic index $\Gamma$
remains constant throughout the
flow.
However, due to the deposition of thermal momentum by outgoing photons, it
is unlikely that $\Gamma$ will remain invariant throughout the
entire trajectory for large-scale astrophysical flows under strong gravity,
even for accretion considered in the steady state, and a modified equation
of state is more realistic, where the adiabatic index is treated as a
function of the spatial distance measured from the gravitational horizon.
Several works in the literature deal with such an equation of
state~\cite{Chandrasekhar1939,Synge1957,RyuChattopadhyayChoi2006,%
Chattopadhyay2008,ChattopadhyayRyu2009,KumarChattopadhyay2014,%
Vyasetal2015,SarkarChattopadhyay2019}, among which the equation of state
proposed in~\cite{RyuChattopadhyayChoi2006,Chattopadhyay2008,%
ChattopadhyayRyu2009} takes a simple form while being applied to study
relativistic transonic flow of multi-component (electrons, ions, and
positrons) accretion onto astrophysical black holes.
Formation of emergent spacetime has been explored for spherically symmetric
accretion of hydrodynamic matter described by the equation of state proposed
by Ryu and collaborators, under the influence of various post-Newtonian
pseudo-Schwarzschild potentials.
In the present work, we will study the formation of the acoustic metric for
general relativistic accretion described by the aforementioned equation of
state onto non-rotating astrophysical black holes.
Michel's work~\cite{Michel1972,ShapiroTeukolsky1983} provided the analysis
of Bondi~\cite{Bondi1952} flow in the Schwarzschild metric; hence the
general relativistic, spherically symmetric astrophysical accretion is
referred to as the Michel flow. Despite
its apparent simplicity, the Michel solution encodes the essential physics of transonic
inflow: the existence of a \emph{critical} or \emph{sonic point} at which the radial
velocity of the fluid equals the local sound speed, and through which a unique, physically
regular solution — the one that passes smoothly from subsonic conditions far from the black
hole to supersonic conditions near the horizon — must thread. This critical-point structure,
and the eigenvalue problem it generates for the accretion rate $\dot{M}$, is the archetype
upon which all more sophisticated models of black-hole accretion are built, and the Michel
solution continues to serve as the indispensable analytic benchmark for general-relativistic
magneto-hydrodynamic codes~\cite{Font2008,Porth2019}.

In our present work, we will first construct and solve the steady-state
equations corresponding to the Michel flow described by the aforementioned
equation of state for a particular flow composition and will obtain the
stationary integral transonic flow solutions in order to present the phase
portrait, where the variation of the flow Mach number will be studied as a
function of the radial distance measured from the black hole event horizon.
We shall then develop a perturbative method to determine the nature of the
flow sonic points.
It will be shown that for non-dissipative spherically symmetric general
relativistic transonic flow, the corresponding critical points will in
general be of saddle type (see,
e.g.,~\cite{Jordan2007,Hilborn2001} for details about the classification
scheme of critical points in the theory of dynamical systems).
The integral accretion solutions will then be perturbed up to the linear
term to obtain the corresponding acoustic spacetime embedded within the
accreting fluid by constructing the acoustic metric that governs the
propagation of the aforementioned linear perturbation.
Such an acoustic metric, which is black hole-like, will be shown to have an
acoustic horizon.
That the acoustic horizon is actually the transonic surface of the flow will
be established by developing the corresponding causal structure through the
construction of the Carter--Penrose diagram.
The analogue version of the surface gravity computed at the acoustic horizon
will then be studied as a function of the accretion parameters.
In this way, the relativistic black hole accretion will be studied from
multiple perspectives --- from the astrophysical perspective, from the point of
view of dynamical systems phenomena, and from the context of emergent gravity
effects.
The basic philosophy behind performing such a multifaceted analysis of
accreting black hole systems has been outlined
in~\cite{GhoseBookChapter2025}.\\

% It is to be noted that every analogue model found in the literature has
% usually been obtained by linearly perturbing transonic (classical or quantum)
% fluids.
% This may suggest that the analogue gravity phenomenon is an artifact of the
% linear perturbation.
% That such an interpretation is not true has very recently been
% proved~\cite{karan_das} by constructing the emergent spacetime through
% the nonlinear perturbation of large-scale astrophysical flow under strong
% gravity, which reinforces the belief that analogue gravity is a far more
% deeply rooted phenomenon in nature than was previously thought.
% In our next work, we will present the nonlinear perturbative analysis of
% the Michel flow with variable $\Gamma$. 

The paper is organized as follows.
Section~\ref{sec:math} presents the general relativistic Michel accretion equations
and derives the critical-point conditions. Section~\ref{sec:dynamic} classifies the transonic solution families. Section~\ref{sec:analogue}
develops the analogue gravity formalism — deriving the acoustic metric from linear
perturbation theory, computing the acoustic surface gravity, and examining its dependence on the energy.
Therein, subsection~\ref{ssec:penrose} presents the Carter-Penrose diagram of the acoustic
spacetime, while subsections~\ref{ssec:surf} and \ref{ssec:riem} discuss the surface gravity and the Riemann curvature tensor for the acoustic spacetime. The stability of the background stationary flow is discussed in section~\ref{sec:stability}. Finally, we briefly summarize our findings and
conclusions in section~\ref{sec:conc}. Throughout the manuscript we use a unit system where $G = c = 1$. The mass of the accretor is also normalized to unity ($M_{BH} = 1$) and radial distances are scaled with Schwarzschild radius ($r_s = 2$).

\section{Mathematical constructions}
\label{sec:math}
We start with an assumption of a general static and spherically symmetric spacetime, whose line element is given by:
\begin{equation}
ds^2=-g_{tt}(r)dt^2+g_{rr}(r)dr^2+r^2(d\theta^2+sin^2\theta d\phi^2),
\label{eq:1}
\end{equation}
where, $r$, $\theta$, and $\phi$ represent the usual spherical polar coordinates.\\

We consider a perfect fluid whose energy momentum tensor $T_{\mu \nu}$  satisfies the below relation :-

\begin{equation}
T_{\mu \nu}=(\epsilon + p)v_{\mu} v_{\nu}+p g_{\mu \nu}
\end{equation}

with $\epsilon$ being the energy density, $p$ pressure and $v_{\mu}$ fluid four velocity satisfying the normalization criteria $v^{\mu}v_{\mu}=-1$.\\

For the fluid flow to be purely radial and inward, the non zero spatial component of velocity $(v^r)$ and the time component $(v^t)$ thus satisfies the relation obtained from normalization :-

\begin{equation}
v^t=\sqrt{\frac{1+g_{rr}v^2}{g_{tt}}}
\label{eq:norm}
\end{equation}

where $v$ represent the radial fluid velocity component, equivalent to $v^{r}$.

\subsection{EoS for a multi-component neutral fluid}
\label{sec:eos}
A longstanding limitation of the hydrodynamic accretion models dealing with spherical accretion is its reliance on a fixed 
adiabatic index $\Gamma$. Real accreting matter traverses a thermodynamic regime that 
is inherently non-uniform: far from the black hole, where temperatures are comparatively 
modest ($kT \ll m_i c^2$), the fluid behaves non-relativistically and $\Gamma \to 5/3$; 
close to the horizon, where temperatures can substantially exceed the rest-mass energy of 
the electrons (and, under certain conditions, of the ions), the fluid is thermally 
relativistic and $\Gamma \to 4/3$. Treating the adiabatic index as a constant throughout 
the flow is therefore a controlled but non-trivial approximation. The exact relativistic 
equation of state (EoS) is expressed in terms of modified Bessel functions of the second 
kind~\cite{Chandrasekhar1939,Synge1957} and, while exact, is computationally and 
analytically unwieldy. 

This difficulty is addressed by the equation of state proposed by 
Chattopadhyay \& Ryu~\cite{ChattopadhyayRyu2009} (hereafter CR~EoS), subsequently 
developed for multispecies relativistic flows by Chattopadhyay and collaborators~\cite{Chattopadhyay2008,RyuChattopadhyayChoi2006}. 
The CR~EoS provides a closed-form, analytic approximation to the Chandrasekhar--Synge 
equation of state that reproduces the exact Bessel-function result to high accuracy across 
the entire range of temperatures, from the non-relativistic to the ultra-relativistic 
regime, while remaining analytically tractable. Crucially, it is formulated for a 
\emph{multispecies} fluid — electrons, positrons, and protons can be treated with 
individual thermal parameters — so that the compositional degree of freedom, absent 
in single-fluid models with a fixed $\Gamma$, is explicitly retained. Physical consequences 
of this compositional freedom are non-trivial: it has been shown, for instance, that 
a pure electron-positron pair plasma is \emph{not} thermally relativistic under 
typical accretion conditions, and that baryon loading is required to drive the effective 
adiabatic index sufficiently below $5/3$ to substantially modify the transonic 
structure~\cite{ChattopadhyayRyu2009}. The CR~EoS has since been employed in a wide 
range of accretion and jet studies~\cite{KumarChattopadhyay2014,Vyasetal2015,SarkarChattopadhyay2019,Joshietal2021}, 
and its role in governing the topology of transonic solutions — including the existence and 
location of shocks — is now well established.
We here consider the fluid to be composed of electrons $(e^-)$, positrons $(e^{+})$ and protons $(p^+)$, where total number density $n$ of the fluid comes from the contributions from individual components and is thus given by:
\begin{equation}
n=n_{e^-}+n_{e^+}+n_{p^+}.
\end{equation}
Propagation of a  fluid in a static and spherically symmetric spacetime with charge neutrality further demands:
\begin{equation} 
n_{e^+}=n_{e^-}(1-\xi),
\end{equation}
where, $\xi=\frac{n_{p^+}}{n_{e^-}}$ represents the relative proportion of protons.
The mass density of the neutral fluid, with $m_i$ being the mass of the individual component is thus given by 
\begin{equation}
\rho = \sum_i n_i m_i=n_{e^-}m_{e^-}\left [2-\xi(1-\frac{1}{\eta}))\right],
\end{equation}
where, $\eta=\frac{m_{e^-}}{m_{p^+}}\approx\frac{1}{1836}$ and the energy density $\epsilon$ being:
\begin{equation}
\epsilon=\sum \left [ n_i m_i c^2 + p_i \left (  \frac{9p_i+3 n_i m_i c^2}{3p_i+2n_im_i c^2} \right) \right]
\label{eq:e}
\end{equation}
We here consider the fluid to be such that its rest mass energy is comparable to its thermal energy $k_{B}T$ with usual sign convention.
Here, we define the dimensionless temperature parameter $\Theta = \frac{k_BT}{m_{e^-} c^2}$ which is a measure of the relative  thermal energy of the fluid at temperature $T$ compared to the electron rest mass energy.
A further simplified form of Eq. (\ref{eq:e}) with the help of the above equations can be written as :-
\begin{equation}
\epsilon=n_{e^-}m_{e^-}c^2f,
\end{equation}
where ,
\begin{equation}
f=(2-\xi)\Bigg[1+\Theta \left(\frac{9\Theta+3}{3\Theta+2}\right)\Bigg]+\xi \Bigg[\frac{1}{\eta}+\Theta\left(\frac{9\Theta+\frac{3}{\eta}}{3\Theta+\frac{2}{\eta}}\right)\Bigg].
\end{equation}
The expression of polytropic index $N$ and effective adiabatic index $\Gamma$ follows from the above and is thus respectively given by:
\begin{equation} 
N=\frac{1}{2}\frac{df}{d\Theta},
\end{equation}
\begin{equation}
\Gamma=1+\frac{1}{N}.
\end{equation}
The definition of sound speed $a^2$ under  adiabatic and isoentropic conditions of this fluid thus becomes:
\begin{equation}
c^2_{\rm s}=\frac{\Gamma p}{p+\epsilon}=\frac{2\Gamma \Theta}{f+2\Theta}.
\label{eq:speed}
\end{equation}\
The more general form of $a^2$ under  isoentropic conditions $(ds=0)$ with pressure $p$ and energy density $\epsilon$ of the fluid is:
\begin{equation}
c^2_{\rm s}=\frac{\partial p}{\partial \epsilon}.
\label{only1}
\end{equation}
Starting from $1^{st}$ law of thermodynamics and for constant entropy $(ds=0)$, the energy density $\epsilon$, mass density $\rho$ and pressure $p$ of the fluid satisfies the condition given by:
\begin{equation}
\frac{\partial \epsilon}{\partial \rho}=\frac{p+\epsilon}{\rho}
\label{only2}
\end{equation}
\subsection{Continuity equation}
Continuity equation ensures mass conservation of the fluid given by $\nabla_{\mu} (\rho v^{\mu})=0$. For radial fluid flow under spherically symmetric conditions simplifies this to:
\begin{equation}
\frac{1}{\sqrt{-g}}\partial_t (\rho v^t \sqrt{-g})+\frac{1}{\sqrt{-g}}\partial_r(\rho v \sqrt{-g})=0,
\label{eq:cont}
\end{equation}
where, $v=v^r$ and $g=det(g_{\mu\nu})=-g_{tt}g_{rr}r^4 sin^2\theta$.
We define $\tilde g=\frac{g}{sin^2\theta}=-g_{tt}g_{rr}r^4$ and for a stationary state of the fluid, Eq.~(\ref{eq:cont}) can be re-written as:
\begin{equation}
\partial_r(\rho v \sqrt{-\tilde g})=0.
\label{eq:cont2}
\end{equation}
Integration of Eq.~(\ref{eq:cont2}) thus gives $\rho_{\rm o} v_{\rm o} \sqrt{-\tilde g}=const$ with subscript $``\rm o"\equiv$ stationary state and for the entire solid angle $\Omega_{\rm solid}$ it becomes, $\Omega_{\rm solid} \,\rho_{\rm o} v_{\rm o} \sqrt{-\tilde g}=const=-\dot M$. Here $\dot M$ is the conserved mass accretion rate or the mass flux with a negative sign representing radially in-falling matter.
Here we define and for the rest of the manuscript, the conserved mass accretion rate  $\psi_{\rm o}(r)$ per unit solid angle as:
\begin{equation}
\psi_{\rm o}(r)=\rho_{\rm o}(r) v_{\rm o}(r) \sqrt{-\tilde g}.
\label{eq:mass1}
\end{equation}
\subsection{Energy momentum conservation equation}
The conservation equation for the energy-momentum tensor of the fluid is given by:
\begin{equation}
\nabla_{\mu} T^{\mu \nu}=0,
\end{equation}
\begin{equation}
\implies \frac{1}{\sqrt{-g}}\partial_\mu(T^{\mu \nu}\sqrt{-g})+\Gamma^{\nu}_{\mu \lambda}T^{\mu \lambda}=0.
\label{eq:19}
\end{equation}
After simplification, Eq. (\ref{eq:19}) can be written as:
\begin{align}
   v^{\mu}\partial_{\mu}v^{\nu}+\frac{v^{\nu}}{\sqrt{-g}}\frac{1}{\epsilon+p}\partial_{\mu}[\sqrt{-g}\frac{\epsilon+p}{\rho}\rho v^{\mu}] & \nonumber \\ +\frac{g^{\mu \nu}}{\epsilon+p}\partial_{\mu}p  +\Gamma_{\mu \lambda}^{\nu}v^{\mu}v^{\lambda} &=0.
\label{eq:20} 
\end{align}

Let the second and third term on the LHS be $T_2$ and  $T_3$ respectively, hence $T_2=\frac{v^{\nu}}{\sqrt{-g}}\frac{1}{\epsilon+p}\partial_{\mu}[\sqrt{-g}\frac{\epsilon+p}{\rho}\rho v^{\mu}]$ and $T_3=\frac{g^{\mu \nu}}{\epsilon+p}\partial_{\mu}p$
Putting the general definition of sound speed (Eq.~(\ref{only1})) and isoentropic conditions of fluid (Eq.~(\ref{only2})), $T_2$ and $T_3$ simplifies to:
\begin{equation*}
T_2=\frac{v^{\mu}v^{\nu}}{\rho}a^2\partial_{\mu}\rho,
\end{equation*}
\begin{equation*}
T_3=\frac{g^{\mu \nu}c^2_{\rm s}}{\rho}\partial_{\mu}\rho.
\end{equation*}
After all the above simplifications, Eq.~(\ref{eq:20}) can be finally re-written as:
\begin{equation}
v^{\mu}\partial_{\mu}v^{\nu}+\frac{v^{\mu}v^{\nu}}{\rho}c^2_{\rm s}\partial_{\mu}\rho+\frac{g^{\mu \nu} c^2_{\rm s}}{\rho}\partial_{\mu}\rho+\Gamma_{\mu \lambda}^{\nu}v^{\mu}v^{\lambda}=0,
\label{eq:21}
\end{equation}
for $\nu=t$, $\nu=r$ in Eq.~(\ref{eq:21}), respectively gives the energy and radial momentum conservation equation. 
Thus energy conservation equation is:
\begin{equation}
v^t \partial_t v^t+v\partial_r v^t+g^{tt}\partial_r g_{tt} v^t v+\frac{a^2}{\rho}[\{(v^t)^2-g^{tt}\}\partial_t \rho+vv^t\partial_r \rho]=0.
\label{eq:energy}
\end{equation}
Integration of time independent part of Eq.~(\ref{eq:energy}) gives:
\begin{equation}
hv_t=-\cal E,
\end{equation}
where, $h$ is the enthalpy of the fluid and $\cal E$ is  relativistic Bernoulli's constant or conserved specific energy along the flow streamlines.
Further, the Euler equation becomes:
\begin{equation}
\begin{split}
&v^t\partial_t v+v\partial_r v+\frac{\partial_r g_{rr}}{2g_{rr}}v^2+\frac{\partial_r g_{tt}}{2g_{rr}}(v^t)^2+\frac{c_{\rm s}^2}{\rho}vv^t\partial_t \rho\\
&+\frac{c^2_{\rm s}}{\rho}(v^2+g^{rr})\partial_r \rho=0.
\label{eq:euler}
\end{split}
\end{equation}
Putting the expression of $v^t$ (Eq.~(\ref{eq:norm})) and under some simplifications, Euler equation (Eq. (\ref{eq:euler})) can be recast as:
% \begin{equation}
% \begin{split}
% &\left(\frac{1+g_{rr}v^2}{g_{tt}}\right)^{1/2}\partial_t v+\frac{c^_{\rm s} v}{\rho}\left(\frac{1+g_{rr}v^2}{g_{tt}}\right)^{1/2}\partial_t \rho\\
% &+\frac{1+g_{rr}v^2}{2g_{rr}}\frac{\partial_r (g_{tt}g_{rr})}{g_{tt}g_{rr}}+\partial_r \left(\frac{1+g_{rr}v^2}{2g_{rr}}\right)\\
% &+\frac{c^_{\rm s}}{\rho}\left(\frac{1+g_{rr}v^2}{2g_{rr}}\right)\partial_r \rho=0.
% \label{eq:euler2}
% \end{split}
% \end{equation}
\begin{align}
&
\left(\frac{1+g_{rr}v^2}{g_{tt}}\right)^{1/2}\partial_t v
+\frac{c^2_{\rm s} v}{\rho}
\left(\frac{1+g_{rr}v^2}{g_{tt}}\right)^{1/2}
\partial_t \rho
\nonumber\\
&\quad
+\frac{1+g_{rr}v^2}{2g_{rr}}
\frac{\partial_r (g_{tt}g_{rr})}{g_{tt}g_{rr}}
+\partial_r\!\left(\frac{1+g_{rr}v^2}{2g_{rr}}\right)
\nonumber\\
&\quad
+\frac{c^2_{\rm s}}{\rho}
\left(\frac{1+g_{rr}v^2}{2g_{rr}}\right)
\partial_r \rho
= 0.
\label{eq:euler2}
\end{align}
Considering only the time independent terms, Eq.~(\ref{eq:euler2}) for a stationary state of the fluid reduces to:
\begin{equation}
-\frac{2c^2_{\rm s}}{\rho}\partial_r \rho=\left(\frac{1+g_{rr}v^2}{2g_{rr}}\right)\partial_r\left(\frac{1+g_{rr}v^2}{2g_{rr}}\right)+\frac{\partial_r (g_{tt}g_{rr})}{g_{tt}g_{rr}}.
\label{eq:euler3}
\end{equation}
Now if $u$ denotes the  fluid radial three velocity in local Lorentz like frame, then $v^t, v, u$ satisfies:
\begin{equation}
v^t=\frac{1}{\sqrt{g_{tt}}}\frac{1}{\sqrt{1-u^2}}, \hspace{2mm} v^2=u^2\frac{1}{g_{rr}(1-u^2)}.
\label{eq:lor}
\end{equation}
For a stationary state of the fluid, considering the radial derivative of conserved mass accretion rate $\psi_{\rm o}(r)$ (Eq.~(\ref{eq:mass1})) in local Lorentz like frame and using the expression in  time independent part of Euler equation (Eq.~( \ref{eq:euler3})) finally gives us:
\begin{equation}
\frac{du}{dr}=\frac{u(1-u^2)}{2(u^2-c^2_{\rm s})}\Big\{c^2_{\rm s} \partial_r \ln{\Big(\frac{-\tilde g}{g_{rr}}\Big)}-\partial_r \ln{g_{tt}} \Big\}.
\label{eq:grad1}
\end{equation}
The expression in Eq.~(\ref{eq:grad1}) clearly shows that denominator vanishes at a radius $r_{c}$ also known as the critical point where $u(r_c)=c_{\rm s}(r_c)$. This  implies available transonic solutions where flow encounters a critical point while transiting from subsonic $(M < 1)$ to supersonic regions $(M > 1)$ towards accretor. Here and for the rest of the manuscript subscript $``c"$ represents critical point where flow velocity equals sound speed.
\begin{figure}
\centering
\includegraphics[width = .7\columnwidth]{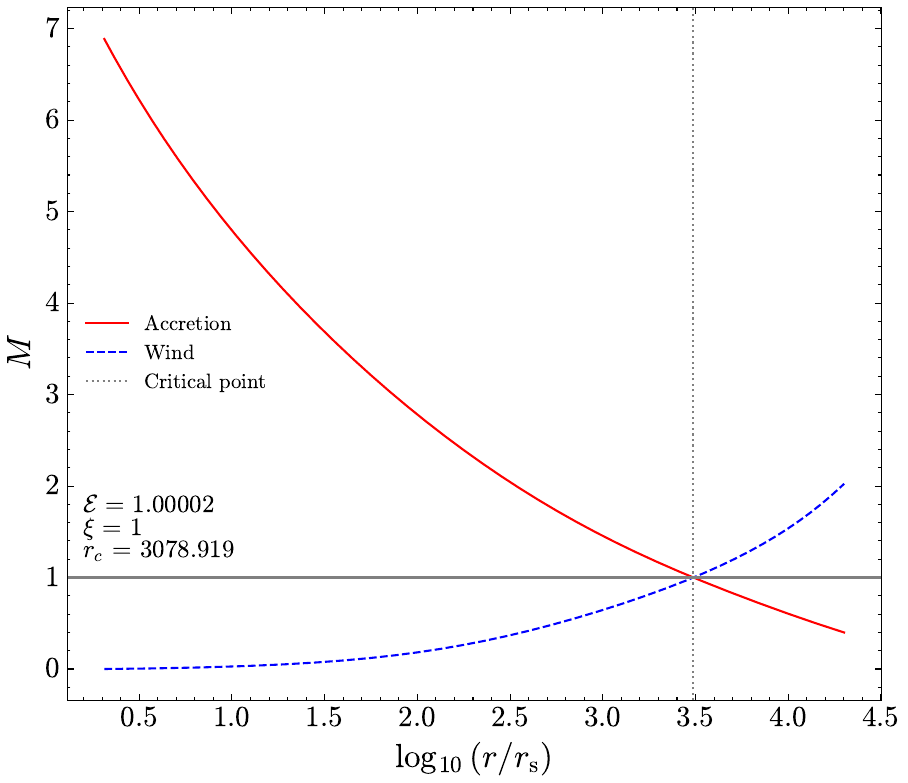}
\caption{Mach number $M$ versus radial distance $log_{10}{r}$ plot. Radial distances are scaled with Schwarzschild radius ($r_{\rm s} = 2$). $\xi = 1.0$ is chosen for the plot.}
\label{fig_mach}
\end{figure}
For a Schwarzschild spacetime metric ($M_{\rm black hole}=G=c=1$): $g_{tt}=g_{rr}^{-1}=(1-\frac{2}{r})$ (Eq.~( \ref{eq:1}) ). Using these in Eq.~(\ref{eq:grad1}) gives us:
\begin{equation} 
\frac{du}{dr}=\frac{(1-u^2) [c^2_{\rm s}(2r-3)-1]}{r(r-2)(u-\frac{c^2_{\rm s}}{u})},
\label{eq:grad2}
\end{equation}
where, analytical expression of sound speed $c^2_{\rm s}$ (Eq.~(\ref{eq:speed})) is determined by EoS of the fluid. It should be noted that in Eq.~(\ref{eq:grad2}), the denominator vanishes at $r=0$, $r=2$ and $r=r_c$. The former two represents the singular points in Schwarzschild spacetime in this coordinate system (Eq.~(\ref{eq:1})), and further we study the phase portrait of the transonic solutions for a stationary state around the critical point $r_c$ only (Fig.~\ref{fig_mach}). A schematic represntation of the flow geometry is found in Fig.~\ref{fig:michel}
At critical point $r_c$, we simultaneously set numerator in Eq.~(\ref{eq:grad2}) to zero thereby ensuring the continuity of flow variable $u(r)$ along each streamlines. This condition further relates $c_{\rm s}\vert_c$ and $r_c$  given by:
\begin{equation}
c^2_{\rm s}\vert_c=\frac{1}{2r_c-3},
\end{equation}

For isentropic fluid flow with EoS (Eq.~(\ref{eq:e})), the  dimensionless temperature gradient in a Schwarzschild spacetime can also be obtained similarly as:

\begin{equation} 
\frac{d\Theta}{dr}=-\frac{\Theta}{N}\left(\frac{2r-3}{r(r-2)}+\frac{1}{u(1-u^2)}\frac{du}{dr}\right)
\label{eq:grad3}
\end{equation}\\

\begin{figure}[t]
\centering
\includegraphics[width = .7\columnwidth]{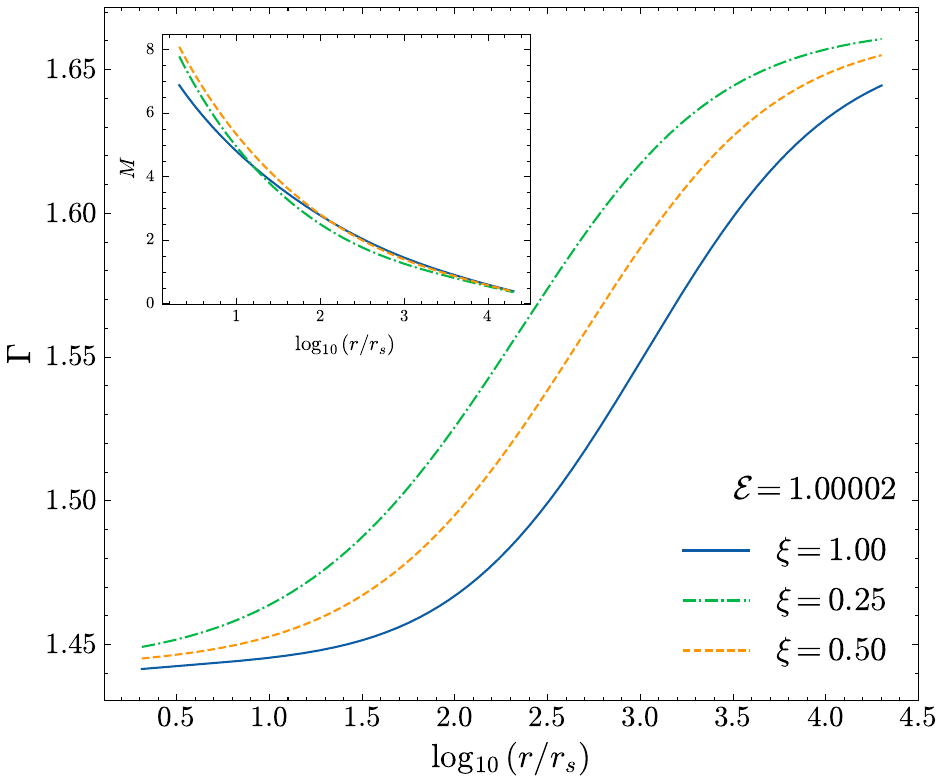}
\caption{Variation of the adiabatic index ($\Gamma$) with radial distance (the corresponding accretion branches are shown in the inset) for different composition of the accreting fluids (i.e. for different $\xi$ values.)}
\label{fig_gamma}
\end{figure}

We further numerically solve the coupled 1st order differential equations Eq.~(\ref{eq:grad2}), Eq.~(\ref{eq:grad3}) involving gradient of $u,\Theta$ using Runge-Kutta method with initial boundary conditions set at $r_c$. A plot of Mach number $M=\frac{u}{c_{\rm s}}$ versus $r$ is shown in Fig.~\ref{fig_mach}. The variation of $\Gamma$ with the radius for this flow is shown in Fig.~\ref{fig_gamma} for different fluid composition (i.e. different $\xi$--values).
\section{Transonic accretion flow from a dynamical system perspective}
\label{sec:dynamic}
\begin{figure}[t]
\centering
\includegraphics[width = .7\columnwidth]{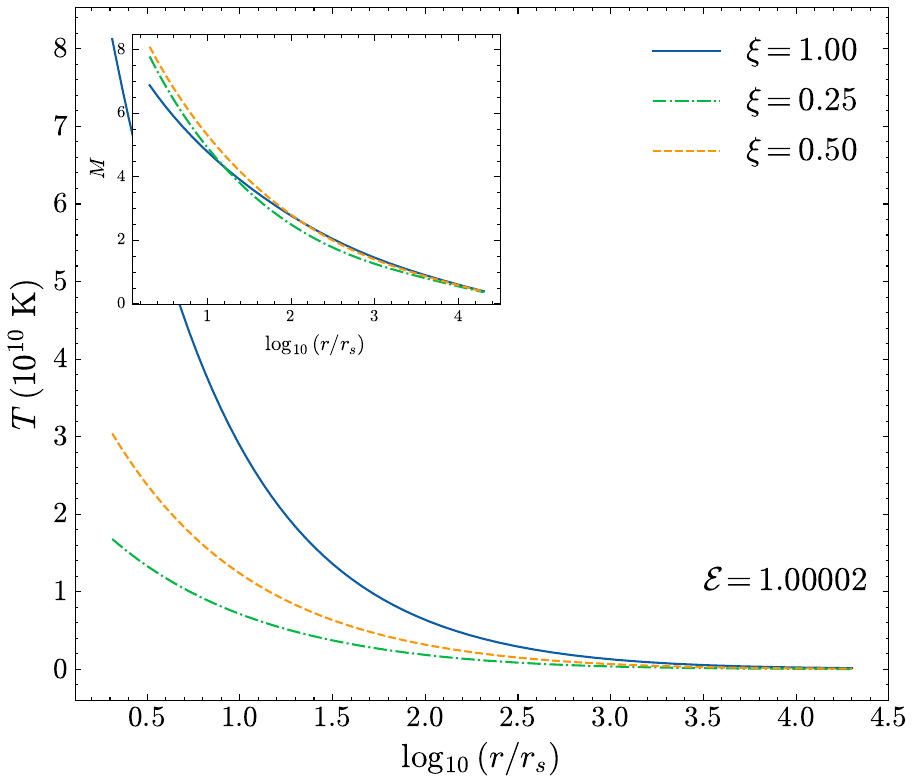}
\caption{Variation of the temperature ($T = \frac{m_ec^2}{k_B}\Theta$) with radial distance (the corresponding accretion branches are shown in the inset) for different composition of the accreting fluids (i.e. for different $\xi$ values.)}
\label{fig_theta}
\end{figure}
For transonic accretion we can introduce a new dynamical variable $\tau$ and write Eq.~(\ref{eq:grad2}) as:
\begin{equation}
\frac{du}{dr}=\frac{\frac{du}{d\tau}}{\frac{dr}{d\tau}}=\frac{N}{D}.
\end{equation}
where, $N=(1-u^2) \{c^2_{\rm s}(2r-3)-1\}$ and $D=r(r-2)(u-\frac{c^2_{\rm s}}{u})$.
Then,
\begin{equation}
\begin{split}
\frac{d}{d\tau}(\delta u) &=\delta N(r,\Theta,u) \implies \frac{\partial N}{\partial r}\delta r+\frac{\partial N}{\partial \Theta}\delta \Theta+\frac{\partial N}{\partial u}\delta u\\
=&2c^2_{\rm s}(1-u^2)\delta r+2c_{\rm s} c_{\rm s}'(1-u^2)(2r-3)\delta \Theta\\
-&2u\{c^2_{\rm s}(2r-3)-1\}\delta u,
\label{eq:53}
\end{split}
\end{equation}
where, $c_{\rm s}' = \frac{dc_{\rm s}}{d\Theta}$. 
\begin{equation}
\begin{split}
\frac{d}{d\tau}(\delta r)&=\delta D(r,\Theta,u)  \implies \frac{\partial D}{\partial r}\delta r+\frac{\partial D}{\partial \Theta}\delta \Theta+\frac{\partial D}{\partial u}\delta u\\
&=(2r-2)(u-\frac{c^2_{\rm s}}{u})\delta r-r(r-2)\Big(\frac{2c_{\rm s}c_{\rm s}'}{u}\Big)\delta \Theta\\
&+r(r-2)(1+\frac{c^2_{\rm s}}{u^2})\delta u.
\label{eq:54}
\end{split}
\end{equation}
$\frac{d\Theta}{dr}$ from Eq.~(\ref{eq:grad3})) satisfying $u(r+\delta r)=u(r)+\delta u$, $c_{\rm s}(r+\delta r)=c_{\rm s}(r)+\delta c_{\rm s}$ is given by:
\begin{equation} 
\delta \Theta=-\frac{\Theta}{N}\Big\{\frac{2r-3}{r(r-2)}\delta r+\frac{1}{u(1-u^2)}\delta u\Big\}.
\end{equation}

Putting $\delta \Theta$ in Eq.~(\ref{eq:53}), Eq.~(\ref{eq:54}) finally yields the form given by:
\begin{equation}
\frac{d}{d\tau}(\delta u)=A_{11}\delta u + A_{12}\delta r,
\label{eq:56}
\end{equation}
\begin{equation}
\frac{d}{d\tau}(\delta r)=A_{21} \delta u + A_{22}\delta r,
\label{eq:57}
\end{equation}
where, coefficients of $\delta u, \delta r$ are:-
\begin{equation}
\begin{split}
A_{11}=&-2c_{\rm s} c_{\rm s}'(2r-3)\frac{\Theta}{N}\frac{1}{u}-2u\{c^2_{\rm s}(2r-3)-1\}\\ 
A_{12}=&2c^2_{\rm s}(1-u^2)-2c_{\rm s} c_{\rm s}'(1-u^2)(2r-3)^2\frac{\Theta}{N}\frac{1}{r(r-2)}\\
A_{21}=&(r-2)(1+\frac{c^2_{\rm s}}{u^2})+r(r-2)\frac{2c_{\rm s}c_{\rm s}' \Theta}{u^2N(1-u^2)}\\
A_{22}=&2c_{\rm s} c_{\rm s}'(2r-3)\frac{\Theta}{N}\frac{1}{u}+(2r-2)(u-\frac{c^2_{\rm s}}{u})\\
\label{eq:59}
\end{split}
\end{equation}
Thus, Eqs.~(\ref{eq:56}) and (\ref{eq:57}) together can be represented in a matrix form as shown below, where $LHS$ represents changes with respect to $\tau$ in the locality of $r$ given by:
\begin{multline}
\begin{pmatrix}
\frac{d}{d \tau}(\delta u)\\
\\
\frac{d}{d \tau}(\delta r)
\end{pmatrix}=
\begin{pmatrix}
A_{11} & A_{12}\\
&\\
A_{21} & A_{22}\\
\end{pmatrix}
\times \begin{pmatrix}
\delta u\\
\\
\delta r
\end{pmatrix}
\label{eq:matrix2}
\end{multline}\\
We further determine the possible trajectory of the transonic solutions in the neighbourhood of $r_c$ as shown below. The above Eq.~(\ref{eq:matrix2}) can be written more compactly as $X'=AX$, where $A$ is $2\times2$ matrix in the $RHS$. If the solution of the above dynamical system of equation are assumed to be $\delta u \sim e^{\Omega(r) \tau}$ and $\delta r \sim e^{\Omega(r)\tau}$, then eigenvalue equation describing this is $AX=\Omega X$. The eigenvalues will be obtained from the characteristic equation given by $det[A-\Omega I]=0$ yielding a quadratic equation:
\begin{figure}[t]
\centering
\includegraphics[width =.8\columnwidth]{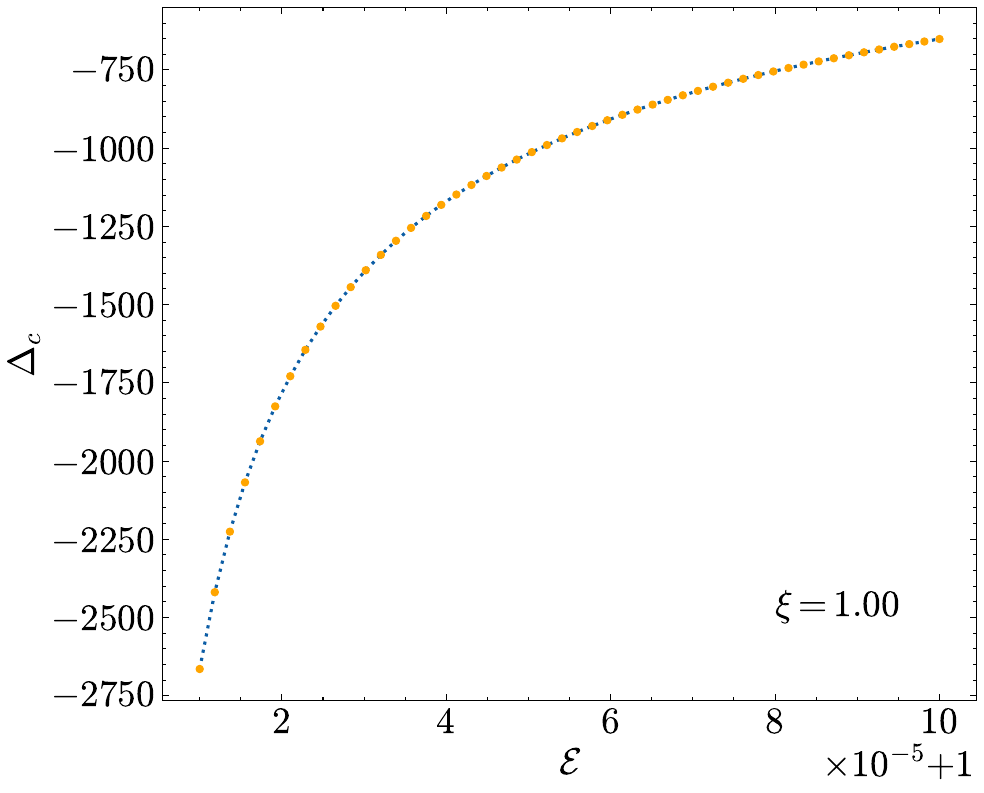}
\caption{Variation of $\Delta_c$ (Eq.~(\ref{eq:det})) for different input specific energies $\cal E$ and $\xi=1$. Here $\Delta_c < 0$ throughout ranging from $\cal E$$=1.00001$ to $1.0001$ implying $\Omega^2_c > 0$ (Eq.~(\ref{eq:62})) and saddle type behaviour for the transonic solutions near the critical point.}
\label{del_e}
\end{figure}
\begin{equation}
\Omega^2(r)+Tr[A]\Omega(r)+\Delta(r)=0,
\label{eq:60}
\end{equation}
where, $\Delta(r)=det[A]$. 
From Eqs.~(\ref{eq:matrix2}) and (\ref{eq:59}), trace $Tr[A](r) = A_{11}+A_{22}=-2u\{c^2_{\rm s}(2r-3)-1\}+(2r-2)(u-\frac{c^2_{\rm s}}{u})$
The trajectories for the transonic solutions locally about $r_c$ is determined by $Tr[A],\Delta$  respectively given by:
\begin{equation}
Tr[A]_{c}=-2u_{c}\{c^2_{\rm s}\vert_c(2r_c-3)-1\}+(2r_c-2)(u_{c}-\frac{c^2_{\rm s}\vert_c}{u_{c}})=0,
\label{eq:trace}
\end{equation}
\begin{equation}
\begin{split}
&\Delta_{c}=-\Big[(r_c-2)\Big(2+\frac{2r_c c'_{\rm s}\vert_c\Theta_c}{N_c}\frac{1}{u_{c}(1-u^2_{c})}\Big)(1-u^2_{c}) \\
&\times \Big(2c^2_{\rm s}\vert_c-\frac{2c'_{\rm s}\vert_c\Theta_c}{c^3_{\rm s}\vert_cN_cr_c(r_c-2)}\Big)+\Big(\frac{2c'_{\rm s}\vert_c\Theta_c}{N_c u_{c} c_{\rm s}\vert_c}\Big)^2\Big].
\label{eq:det}
\end{split}
\end{equation}
Therefore from Eq.~(\ref{eq:60}), $\Omega$ for the critical point is
\begin{equation}
\Omega^2_{c}=-\Delta_{c}
\label{eq:62}
\end{equation}
Clearly the nature of roots $\Omega$ will depend on $Tr[A]$ and discriminant $Tr[A]^2-4\Delta$ (Eq.~(\ref{eq:60})). 
From Eq.~(\ref{eq:59}), Eq.~(\ref{eq:matrix2}) it is clear that $A$ is expressed in terms of the flow variables at different radii and $Tr[A]=0$ for the critical point (Eq.~(\ref{eq:trace})). Further $\Delta < 0$ for the critical point  will yield two real opposite values of $\Omega$ (Eq.~(\ref{eq:62})). The real, opposite $\Omega's$ for critical point corresponding to different $\cal E$  (Fig.~\ref{del_e}) ensures saddle type behaviour for the transonic solutions near the critical point, similar to as shown in Fig.~\ref{fig_mach} and consistent for a non-viscous fluid flow.
\begin{figure}
\centering
\includegraphics[width = .9\columnwidth]{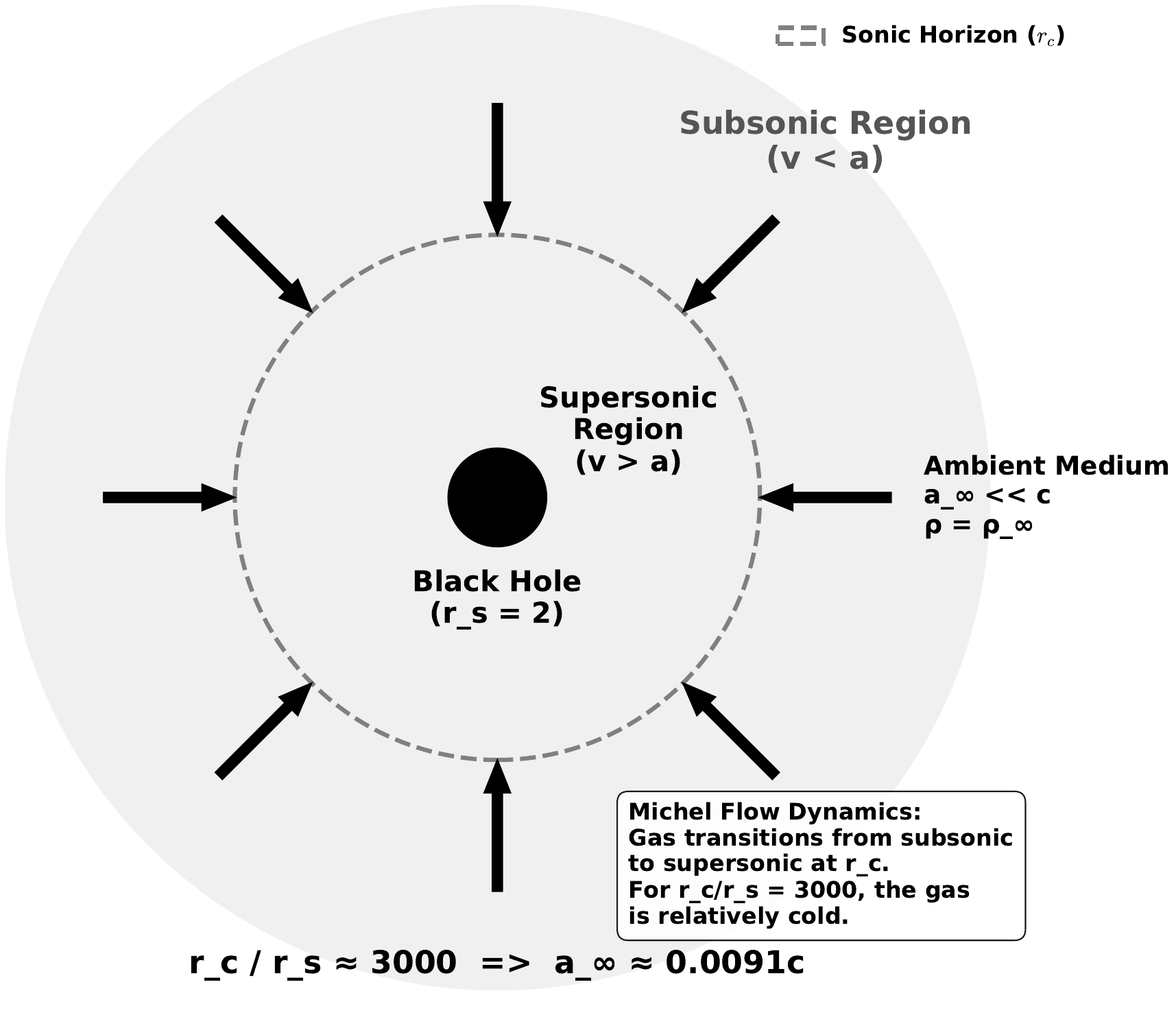}
\caption{Schematic representation of the accretion flow depicted in Fig.~\ref{fig_mach}. }
\label{fig:michel}
\end{figure}
\section{Perturbation of mass accretion rate and emergence of acoustic geometry}
\label{sec:analogue}
From the expression of the conserved mass accretion rate (Eq.~(\ref{eq:mass1})), $\rho_{\rm o}$ is a
scalar, $v_{\rm o}$ is the radial component of the four-velocity, and $\tilde{g}$ is related to the
background spacetime metric. Consequently, unlike the velocity potential $\phi$, the quantity
$\psi_{\rm o}$ will not transform as a scalar under a general coordinate transformation.
For an irrotational fluid flow, the four-velocity and the velocity potential $\phi$ are
related through $h v_{\mu} = \nabla_{\mu}\phi$, where $h$ is the specific enthalpy.
Substituting this relation into Eq.~(\ref{eq:mass1}) yields
\begin{equation}
    \psi_{\rm o} = \frac{\rho_{\rm o} \sqrt{-\tilde{g}}\,\partial_r \phi}{h}.
\end{equation}
We now define the quantity $\psi_{\rm} = \rho v \sqrt{-\tilde{g}}$, whose spatially conserved
stationary state coincides with $\psi_{\rm o}$ (Eq.~(\ref{eq:mass1})).
We employ a time-dependent linear perturbation scheme by writing
\begin{align}
    v(r,t) &= v_{\rm o}(r) + \epsilon\, v'(r,t), \nonumber \\
    \rho(r,t) &= \rho_{\rm o}(r) + \epsilon\, \rho'(r,t), \nonumber \\
    \psi(r,t) &= \psi_{\rm o}(r) + \epsilon\,\psi'(r,t), \nonumber\\
    c_{\rm s}(r,t) &= c_{\rm so}(r) + \epsilon\,c_{\rm s}'(r,t),
\end{align}
where $\epsilon \ll 1$. Note that the $r$ dependence of $c_{\rm s}$ comes from the fact that $c_{\rm s}(\Theta)$ and $\Theta = \Theta(r)$. For the rest of the manuscript any quantity with a subscript `${\rm o}$' would relate to the background (stationary) flow.  Substituting into $\psi$ and retaining terms to first order in
$\epsilon$ gives
\begin{equation}
    \psi'(r,t) = \bigl(\rho'\,v_o + \rho_o\,v'\bigr)\sqrt{-\tilde{g}}.
    \label{eq:psi1}
\end{equation}
Substituting the perturbed $\rho$ and $v$ into the continuity equation
(Eq.~(\ref{eq:cont})) and collecting orders in $\epsilon$, the zeroth-order term gives
\begin{equation}
    \frac{1}{\sqrt{-\tilde{g}}}\,\partial_t\!\left(\rho_{\rm o} v_{\rm o}^t \sqrt{-\tilde{g}}\right)
    + \frac{1}{\sqrt{-\tilde{g}}}\,\partial_r\!\left(\sqrt{-\tilde{g}}\,\rho_{\rm o} v_{\rm o}\right) = 0,
\end{equation}
which is simply the stationary continuity equation, while the first-order term yields
\begin{align}
    \frac{1}{\sqrt{-\tilde{g}}}\,\partial_t\!\left[
        \left(\rho' v_{\rm o}^t + \frac{v_{\rm o} v' g_{rr}\rho_{\rm o}}{g_{tt}v^t}\right)\sqrt{-\tilde{g}}
    \right] \nonumber \\
    + \frac{1}{\sqrt{-\tilde{g}}}\,\partial_r\!\left[
        \left(\rho_{\rm o} v' + \rho' v_{\rm o}\right)\sqrt{-\tilde{g}}
    \right] = 0,
\end{align}
which, expressed in terms of $\psi'$ (Eq.~(\ref{eq:psi1})) and
$v_{\rm o}^t = \sqrt{(1 + g_{rr}v_{\rm o}^2)/g_{tt}}$, becomes
\begin{equation}
    \frac{1}{\sqrt{-\tilde{g}}}\,\partial_t\!\left[
        \left(\rho' v_{\rm o}^t + \frac{v_{\rm o} v' g_{rr}\rho_{\rm o}}{g_{tt}v_{\rm o}^t}\right)\sqrt{-\tilde{g}}
    \right]
    + \frac{1}{\sqrt{-\tilde{g}}}\,\partial_r\psi' = 0.
    \label{eq:36}
\end{equation}

Taking the time derivative of $\psi'$ (Eq.~(\ref{eq:psi1})) and substituting into
Eq.~(\ref{eq:36}), one obtains the perturbed velocity and density in terms of $\psi'$:
\begin{equation}
    \partial_t v' = \frac{g_{tt}v_{\rm o}^t}{\rho_{\rm o}\sqrt{-\tilde{g}}}
    \left(v_{\rm o}^t\,\partial_t\psi' + v_{\rm o}\,\partial_r\psi'\right),
\end{equation}
\begin{equation}
    \partial_t\rho' = -\frac{1}{\sqrt{-\tilde{g}}}
    \left(g_{rr}v_{\rm o}\,\partial_t\psi' + g_{tt}v_{\rm o}^t\,\partial_r\psi'\right).
\end{equation}

Substituting the perturbed variables into the Euler equation (Eq.~(\ref{eq:euler2})) and
invoking the stationary background conditions yields
\begin{equation}
    \frac{g_{rr}}{g_{tt}v_{\rm o}^t}\,\partial_t v'
    + \partial_r\!\left(\frac{g_{rr}v_{\rm o} v'}{g_{tt}(v_{\rm o}^t)^2}
    + \frac{c_{\rm s}^2\,\rho'}{\rho_{\rm o}}\right)
    + \frac{g_{rr}v_{\rm o} c_{\rm s}^2}{g_{tt}\rho_{\rm o} v_{\rm o}^t}\,\partial_t\rho' = 0.
\end{equation}

Taking the time derivative of this equation and substituting the expressions for
$\partial_t\rho'$ and $\partial_t v'$ derived above, one finally arrives at the
space-time propagation equation for the linearly perturbed mass accretion rate $\psi'$:
\begin{equation}
    \partial_{\mu}\!\left(f^{\mu\nu}\,\partial_{\nu}\psi'\right) = 0,
    \label{eq:alemb}
\end{equation}
where $\mu,\nu \in \{r,t\}$ and $f^{\mu\nu}$ is the symmetric $2\times 2$ matrix
% \begin{equation}
%     f^{\mu\nu} = \frac{g_{rr}v c_{\rm s}^2}{v^t}
%     \begin{pmatrix}
%         -g^{tt} + \!\left(1-\dfrac{1}{c_{\rm s}^2}\right)(v^t)^2
%         & v v^t\!\left(1-\dfrac{1}{c_{\rm s}^2}\right) \\[10pt]
%         v v^t\!\left(1-\dfrac{1}{c_{\rm s}^2}\right)
%         & g^{rr} + \!\left(1-\dfrac{1}{c_{\rm s}^2}\right)v^2
%     \end{pmatrix}.
%     \label{eq:matrix}
% \end{equation}
\begin{subequations}
\label{eq:matrix}
\begin{align}
    f^{\mu\nu} &= \frac{g_{rr}v_{\rm o} c^2_{\rm so}}{v_{\rm o}^t}\,\tilde{f}^{\mu\nu}, 
    \qquad \beta \equiv 1 - c^2_{\rm so}, \label{eq:matrixa}\\
    \tilde{f}^{tt} &= -g^{tt} + \beta\,(v_{\rm o}^t)^2, \label{eq:matrixb}\\
    \tilde{f}^{rr} &= g^{rr} + \beta\, v_{\rm o}^2, \label{eq:matrixc}\\
    \tilde{f}^{tr} &= \tilde{f}^{rt} = \beta\, v_{\rm o} v_{\rm o}^t. \label{eq:matrixd}
\end{align}
\end{subequations}
The structural similarity between Eq.~(\ref{eq:alemb}) and the covariant wave equation for a
massless scalar field in curved spacetime is the central observation that establishes the
analogue gravity correspondence. In a spacetime with metric $G_{\mu\nu}$ and determinant
$G = \det(G_{\mu\nu})$, the massless Klein--Gordon equation reads~\cite{Wald1984,
BarceloLiberatiVisser2005}
\begin{equation}
    \Box_{G}\,\Psi \equiv
    \frac{1}{\sqrt{-G}}\,\partial_{\mu}\!\left(\sqrt{-G}\,G^{\mu\nu}\,\partial_{\nu}\Psi\right) = 0.
    \label{eq:KG}
\end{equation}
Equation~\ref{eq:alemb} is identically of this form provided one identifies
\begin{equation}
    \sqrt{-G}\,G^{\mu\nu} \;\equiv\; f^{\mu\nu},
    \label{eq:identify}
\end{equation}
so that the perturbed accretion variable $\psi'$ plays the role of the massless scalar
field $\Psi$, and the matrix $f^{\mu\nu}$ encodes the effective (acoustic) spacetime
geometry~\cite{Unruh1981,Visser1998,Das2004}.

\paragraph{Contravariant acoustic metric.}
To extract the contravariant components $G^{\mu\nu}$ from Eq.~(\ref{eq:identify}) we
first compute the determinant of $f^{\mu\nu}$:
\begin{align}
    \det(f^{\mu\nu}) &= \mathcal{F}^2
    \Bigl[\bigl(-g^{tt}+\beta\,(v_{\rm o}^t)^2\bigr)
          \bigl(g^{rr}+\beta\, v_{\rm o}^2\bigr) \nonumber\\
    &\qquad\qquad -\,\beta^2\, v_{\rm o}^2(v_{\rm o}^t)^2
    \Bigr].
    \label{eq:detf}
\end{align}
where, $\mathcal{F} = g_{rr}v_{\rm o} c^2_{\rm so}/v^t$.
Since $\det(\sqrt{-G}\,G^{\mu\nu}) = (-G)\cdot\det(G^{\mu\nu}) = (-G)/G = -1$ in
two dimensions (the $r$-$t$ sector), one has $\det(f^{\mu\nu}) = -1$, which provides a
consistency condition on the background flow variables. From the identification
$\sqrt{-G} = \sqrt{-\det(f^{\mu\nu})^{-1}\,[\det(f)]^2}$, one obtains the overall
conformal factor
\begin{equation}
    \sqrt{-G} = \left(-\det f^{\mu\nu}\right)^{1/2},
\end{equation}
and the contravariant acoustic metric components follow as
\begin{equation}
    {
    G^{\mu\nu} = \frac{f^{\mu\nu}}{\sqrt{-\det(f^{\alpha\beta})}}.
    }
    \label{eq:Gup}
\end{equation}
Explicitly, identifying $f^{\mu\nu}$ with Eq.~(\ref{eq:matrix}):
\begin{equation}
    G^{\mu\nu} = \mathcal{N}
    \begin{pmatrix}
        -g^{tt}+\beta\,(v_{\rm o}^t)^2
        & \beta\, v_{\rm o}v_{\rm o}^t\\[6pt]
        \beta\, v_{\rm o}v_{\rm o}^t
        & g^{rr}+\beta\, v_{\rm o}^2
    \end{pmatrix},
    \label{eq:Gcontra}
\end{equation}
where $\mathcal{N} = \mathcal{F}/\sqrt{-\det(f^{\alpha\beta})}$ is a scalar conformal factor
that depends on the background flow variables and the equation of state through $c_{\rm s}$.

\paragraph{Covariant acoustic metric.}
The covariant components $G_{\mu\nu}$ are obtained by matrix inversion of
$G^{\mu\nu}$. For a $2\times 2$ Lorentzian matrix with components
$G^{\mu\nu} = \mathcal{N}\,\tilde{f}^{\mu\nu}$, the inverse is
\begin{equation}
    G_{\mu\nu} = \frac{1}{\mathcal{N}\,\det(\tilde{f}^{\mu\nu})}
    \begin{pmatrix}
         \tilde{f}^{rr} & -\tilde{f}^{tr}\\[4pt]
        -\tilde{f}^{rt} &  \tilde{f}^{tt}
    \end{pmatrix},
\end{equation}
where $\tilde{f}^{\mu\nu}$ denotes the matrix inside the parentheses in
Eq.~(\ref{eq:Gcontra}). Carrying out the inversion explicitly gives the covariant
acoustic metric~\cite{Visser1998,ShaikhDas2018}
\begin{equation}
    G_{\mu\nu} = \frac{1}{\mathcal{N}(-\det \tilde{f})}
    \begin{pmatrix}
        g^{rr}+\beta\, v_{\rm o}^2
        & -\beta\, v_{\rm o}v_{\rm o}^t\\[6pt]
        -\beta\, v_{\rm o}v_{\rm o}^t
        & -g^{tt}+\beta\,(v_{\rm o}^t)^2
    \end{pmatrix}.
    \label{eq:Gcov}
\end{equation}
It is straightforward to verify that $G^{\mu\alpha}G_{\alpha\nu} = \delta^{\mu}{}_{\nu}$,
confirming the consistency of the inversion.

\paragraph{Acoustic line element.}
The acoustic spacetime interval is defined through the covariant metric (Eq.~(\ref{eq:Gcov}))
in the standard way. Written out explicitly, the line element in the effective acoustic
spacetime reads
\begin{align}
    \label{eq:lineel}
    dS^2 & = G_{\mu\nu}\,dx^{\mu}dx^{\nu}
    = \Omega^2 \nonumber \\ 
    & = \left(g^{rr}+\beta v^2\right)dt^2 - 2\beta vv^t\,dt\,dr \nonumber \\
& \qquad \qquad + \left(-g^{tt}+\beta(v^t)^2\right)dr^2,
\end{align}
where $\Omega^2$ is an overall conformal factor (a function of the background flow
variables) whose precise form is set by the normalization in Eq.~(\ref{eq:Gcov}). The
structure of Eq.~(\ref{eq:lineel}) is the canonical form of the acoustic metric first
identified by Unruh~\cite{Unruh1981} and systematically developed by
Visser~\cite{Visser1998}: it is a Painlev\'e--Gullstrand-type metric \cite{Painleve1921, Gullstrand1922} in which the
effective ``river'' velocity is $v(r)$ and the local ``speed of light'' is replaced
by the adiabatic sound speed $c_{\rm so}(r)$.

The acoustic horizon — the analogue of the event horizon — is located at the radial
position $r = r_c$, where,
\begin{equation}
    c^2_{\rm s}\vert_c = v^2_c,
    \label{eq:horizon}
\end{equation}
i.e., where the infall velocity of the fluid equals the local sound speed (see Fig.~\ref{fig:michel} for a scematic representation of the sonic surface). This is
precisely the sonic (critical) point (surface) of the stationary background flow. At $r = r_c$,
the coefficient of $dt^2$ in Eq.~(\ref{eq:lineel}) vanishes: the acoustic metric is
degenerate there in the same sense that the Schwarzschild metric is degenerate at the
event horizon in Schwarzschild coordinates, signaling the presence of an acoustic
horizon across which acoustic perturbations cannot propagate outward.  It should be noted that Eq.~(\ref{eq:alemb}), and consequently the acoustic metric
(Eqs.~(\ref{eq:Gcontra})--(\ref{eq:lineel})), is valid for any neutral fluid under adiabatic
and isentropic conditions flowing in a general static, spherically symmetric spacetime of
the form given by Eq.~(\ref{eq:1})~\cite{ShaikhDas2018,BarceloLiberatiVisser2005}. The
composition-dependence of the acoustic geometry enters entirely through the sound speed
$c^2_{\rm s}$, whose analytical form is determined by the equation of state of the fluid
(Eq.~(\ref{eq:speed})). In the present work, this is the CR~EoS of
Chattopadhyay \& Ryu~\cite{ChattopadhyayRyu2009}, which makes the acoustic geometry —
and in particular the surface gravity $\kappa$ — explicit functions of the fluid composition and the conserved
specific energy $\mathcal{E}$ of the background Michel flow.
\subsection{Causal Structure of The Emergent Space-Time}
\label{ssec:penrose}
\begin{figure}[t]
\includegraphics[width=.8\columnwidth]{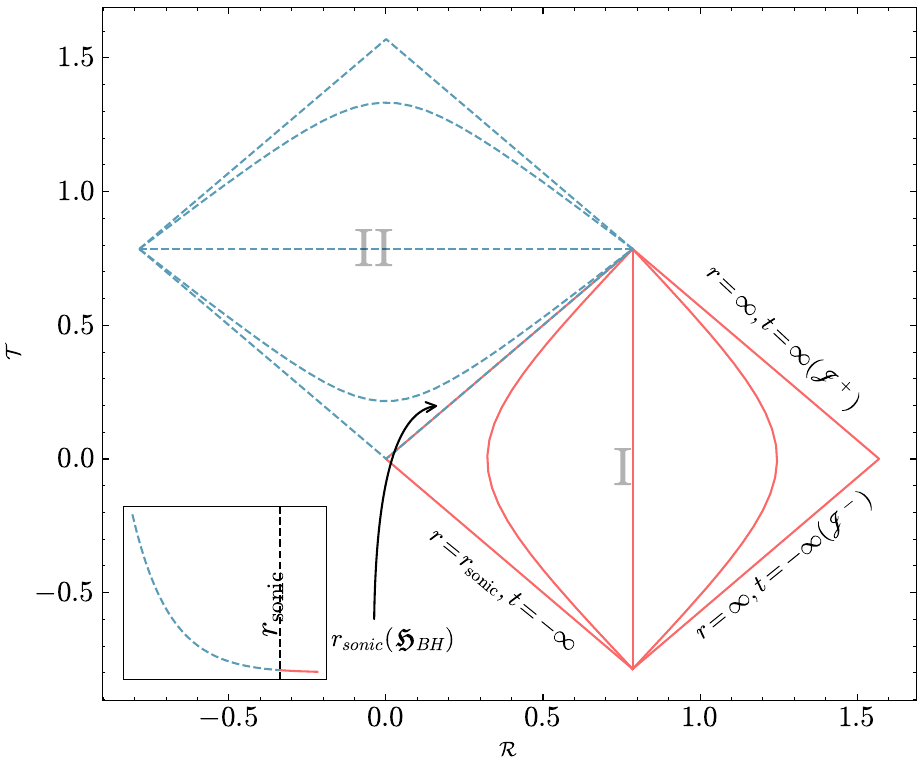}
\caption{Carter--Penrose diagram of the acoustic spacetime for the transonic Michel
flow of Fig.~\ref{fig_mach}. Region~I ($r > r_{\rm sonic}$, red) and Region~II
($r < r_{\rm sonic}$, blue dashed) are separated by the acoustic horizon
$\mathfrak{H}_{\rm BH}$ at $r = r_{\rm sonic}$. Future and past null infinities
$\mathscr{I}^{\pm}$ bound Region~I at $r \to \infty$. The physical accretion branch
is shown in the inset.}
\label{fig:pcd}
\end{figure}

\begin{figure}[t]
\includegraphics[width=.8\columnwidth]{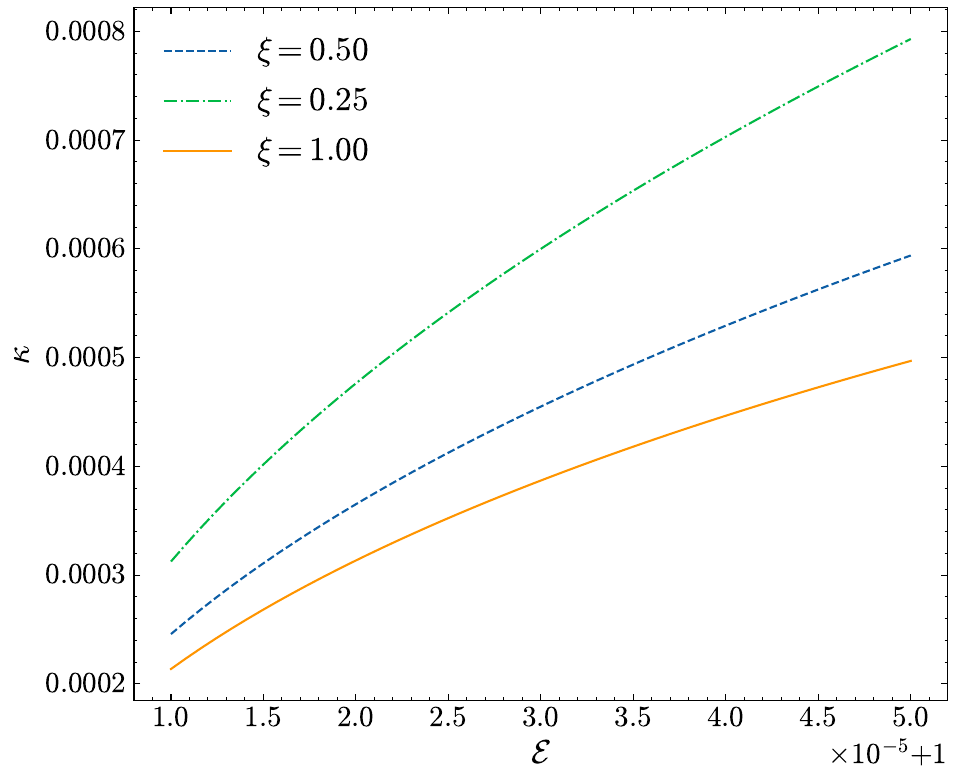}
\caption{Acoustic surface gravity $\kappa$ as a function of the conserved specific
energy $\mathcal{E}$ for a different fluid compositions (i.e. different $\xi$). Since $\mathcal{E}$
uniquely fixes the sonic radius via the critical-point conditions, the plot also
describes the dependence of $\kappa$ on the acoustic horizon location.}
\label{fig:kappa}
\end{figure}

The global causal structure of the acoustic spacetime is encoded in the Carter--Penrose
diagram (Fig.~\ref{fig:pcd}), constructed via the standard sequence of tortoise,
Eddington--Finkelstein, Kruskal--Szekeres, and conformal compactification coordinate
transformations applied to $G_{\mu\nu}$~\cite{HawkingEllis1973,Wald1984,Visser1998,
ShaikhDas2018}. For the detailed description of the Carter--Penrose coordinates in similar acoustic space-time context please the reader is advised to see sec. VI, sub-section A of \cite{GhoseDas2025}. The diagram displays two causally distinct regions separated by the
acoustic horizon $\mathfrak{H}_{\rm BH}$ at $r = r_{\rm sonic}$. Region~I
($r > r_{\rm sonic}$) is the exterior domain bounded by $\mathscr{I}^{\pm}$ at
$r \to \infty$. For an accretion flow, this is the subsonic region. Region~II ($r < r_{\rm sonic}$) is the acoustic black hole interior (supersonic region for accretion)
from which no perturbation can propagate outward, in exact analogy with the Schwarzschild
event horizon. The full two-diamond structure is the maximal analytic extension of the
acoustic metric~\cite{refId0}; the physical one-way accretion solution
occupies Region~I and the future horizon, as shown in the inset.
\subsection{Acoustic surface gravity}
\label{ssec:surf}
\begin{figure}[t]
\includegraphics[width=.8\columnwidth]{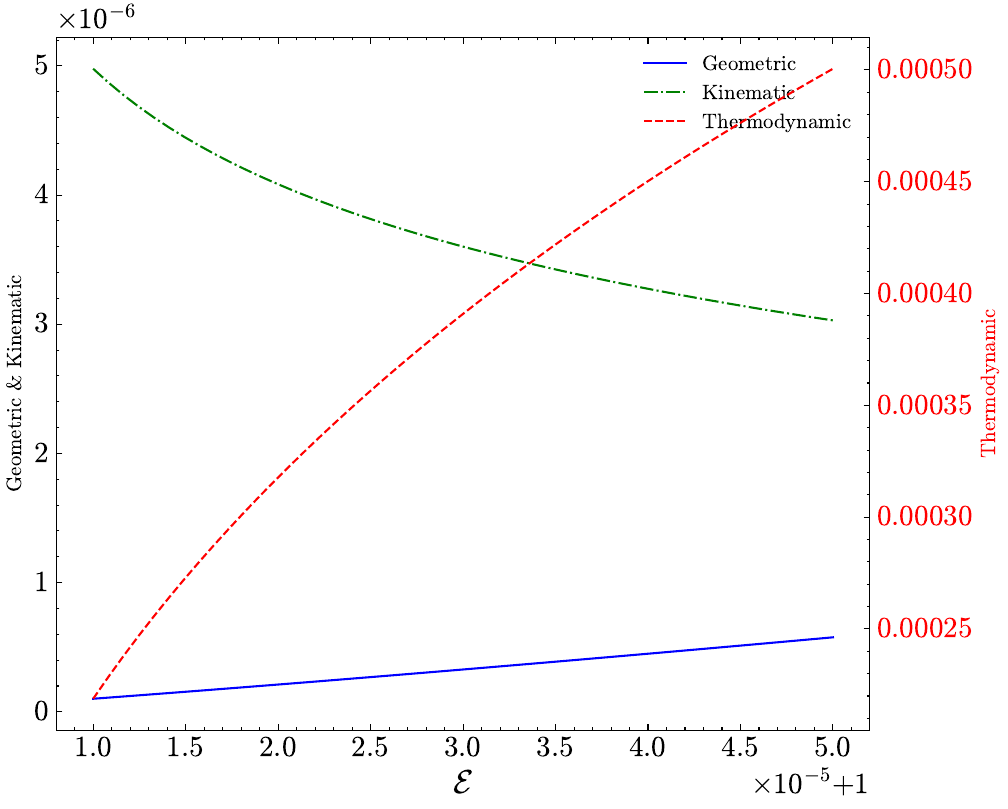}
\caption{As seen from Eq.~(\ref{eq:kappa}), geometry, thermodynamics, and kinematics all contribute to the surface gravity. The above plot compares these contributions for different accretion energy, corresponding to different critical radii (acoustic horizons at different radial distance).}
\label{fig:kcomp}
\end{figure}

The acoustic metric (Eq.~(\ref{eq:Gcov})) is stationary, admitting the timelike Killing
vector $\xi^{\mu} = (\partial_t)^{\mu} = (1,0)$. The acoustic horizon
$\mathfrak{H}_{\rm BH}$ is the Killing horizon of $\xi^\mu$, defined as the null
surface where
\begin{equation}
    G_{\mu\nu}\,\xi^{\mu}\xi^{\nu} = G_{tt} = g_{rr} + \beta\, v^2 = 0,
    \label{eq:killing_null}
\end{equation}
which at the sonic point $v_c = a_c$ gives the critical point condition
\begin{equation}
    \beta_c = -\frac{g_{rr}(r_c)}{c_{\rm s}\vert_c^2}
    = -\frac{1}{\mathcal{A}_c\,c_{\rm s}\vert_c^2}, \qquad
    \mathcal{A}_c \equiv 1-\frac{r_s}{r_c}.
    \label{eq:horizon_cond}
\end{equation}
The surface gravity $\kappa$ associated with this Killing horizon is defined
through~\cite{Wald1984} (in the context of acoustic space-time see Sec.~6 of
\cite{Abraham:2005ah})
\begin{equation}
    \nabla_\mu\!\left(G_{\alpha\beta}\,\xi^\alpha\xi^\beta\right)\big|_{\mathfrak{H}_{\rm BH}}
    = -2\kappa\,G_{\mu\nu}\,\xi^\nu\big|_{\mathfrak{H}_{\rm BH}},
    \label{eq:kappa_killing}
\end{equation}
which for a metric of the form of Eq.~(\ref{eq:Gcov}) reduces to
\begin{equation}
    \kappa
    = \frac{\left|\partial_r G_{tt}\right|_{r_c}}{2\left|G_{tr}\right|_{r_c}}
    = \frac{\left|\partial_r G_{tt}\right|_{r_c}}
    {2\,|\beta_c|\,c_{\rm s}\vert_c\,v_c^{t}},
    \label{eq:kappa_def}
\end{equation}
where we used $|G_{tr}|_{r_c} = |\beta_c|\,a_c\,v_c^{t}$ at the horizon.
Expanding $\partial_r G_{tt}$ explicitly,
\begin{equation}
    \partial_r G_{tt}\big|_{r_c}
    =
    \underbrace{\frac{r_s}{r_c^2\,\mathcal{A}_c^2}}_{\rm geometric}
    +
    \underbrace{\frac{1}{c_{\rm s}\vert_c^2}
    \left(\frac{d c_{\rm s}^2}{d\Theta}\right)_c
    \left.\frac{d\Theta}{dr}\right|_{r_c}}_{\rm thermodynamic}
    +
    \underbrace{2\beta_c\,c_{\rm s}\vert_c
    \left.\frac{dv}{dr}\right|_{r_c}}_{\rm kinematic},
    \label{eq:dGtt}
\end{equation}
where the thermodynamic term follows from
$d\beta/dr = (1/c_{\rm s}^4)\,(d c_{\rm s}^2/d\Theta)\,(d\Theta/dr)$ using the CR~EoS.
The final expression for the acoustic surface gravity is therefore
\begin{equation}
    \kappa
    =
    \frac{1}{2\,|\beta_c|\,c_{\rm s}\vert_c\,v_c^{t}}
    \left|
    \frac{r_s}{r_c^2\,\mathcal{A}_c^2}
    +
    \frac{1}{c_{\rm s}\vert_c^2}
    \left(\frac{d c_{\rm s}^2}{d\Theta}\right)_c
    \left.\frac{d\Theta}{dr}\right|_{r_c}
    +
    2\beta_c\,c_{\rm s}\vert_c
    \left.\frac{dv}{dr}\right|_{r_c}
    \right|.
    \label{eq:kappa}
\end{equation}
The variation of $\kappa$ with $\mathcal{E}$ is shown in Fig.~\ref{fig:kappa} for different fluid compositions (i.e. for different values of $\xi$). The surface gravity increases monotonically with $\mathcal{E}$: higher-energy
flows place the sonic point closer to the gravitational horizon where flow gradients are
steeper, yielding larger $\kappa$ through Eq.~(\ref{eq:kappa}). The same equation also shows that the background geometry, thermodynamics, and the kinematics of the flow all contributes to the acoustic surface gravity. A comparison of these contributions are shown in Fig.~\ref{fig:kcomp}. As expected, the most significant contribution to the acoustic surface gravity comes from the thermodynamics of the flow.
\subsection{Riemann curvature tensor for acoustic spacetime}
\label{ssec:riem}
We now calculate the components of Riemann curvature tensor $[R^\rho_{\sigma \mu \nu}]$ for the emergent acoustic spacetime represented by $G_{\mu \nu}$ (Eq \ref{eq:Gcov}). The general form is given by:
\begin{equation}
R^{\rho}_{\sigma \mu \nu}=\partial_{\mu}\Gamma^{\rho}_{\nu\sigma}-\partial_{\nu}\Gamma^{\rho}_{\mu\sigma}+\Gamma^{\rho}_{\mu\lambda}\Gamma^{\lambda}_{\nu\sigma}-\Gamma^{\rho}_{\nu\lambda}\Gamma^{\lambda}_{\mu\sigma},
\end{equation}
where, $\{\rho,\sigma,\mu,\nu\} \in r,t$. The acoustic metric components $G_{\mu \nu}$ are time independent and radially dependent (Eq.~(\ref{eq:Gcov})), thus the Christoffel  symbols corresponding to this metric are given by:
\begin{equation}
\begin{split}
\Gamma^t_{tt}&=-\frac{1}{2}G^{tr}\partial_r G_{tt}\\
\Gamma^t_{tr}=\Gamma^t_{rt}&=\frac{1}{2}G^{tt}\partial_r G_{tt}\\
\Gamma^t_{rr}&=G^{tt}\partial_r G_{tr}+\frac{1}{2}G^{tr}\partial_r G_{rr}\\
\Gamma^r_{rt}=\Gamma^r_{tr}&=\frac{1}{2}G^{rt}\partial_r G_{tt}\\
\Gamma^r_{rr}&=\frac{1}{2}G^{rr}\partial_r G_{rr}+G^{rt}\partial_r G_{tr}\\
\Gamma^r_{tt}&=-\frac{1}{2}G^{rr}\partial_r G_{tt}\\
\end{split}
\label{eq:cris}
\end{equation}
The tensor components are:
\begin{equation*}
\begin{split}
R^{\rho}_{\sigma tr}=&-R^{\rho}_{\sigma r t}=-\partial_r \Gamma^{\rho}_{t\sigma}+\Gamma^{\rho}_{t\lambda}\Gamma^{\lambda}_{r\sigma}-\Gamma^{\rho}_{r \lambda}\Gamma^{\lambda}_{t\sigma}\\
R^{t}_{rtr}=&-R^t_{rrt}=-\partial_r\Gamma^t_{tr}+\Gamma^t_{t\lambda}\Gamma^{\lambda}_{rr}-\Gamma^t_{r\lambda}\Gamma^{\lambda}_{tr}\\
\implies& -\partial_r\Gamma^t_{tr}+\Gamma^t_{tt}\Gamma^t_{rr}+\Gamma^t_{tr}\Gamma^r_{rr}-\Gamma^t_{rt}\Gamma^t_{tr}-
\Gamma^t_{rr}\Gamma^r_{tr}.
\end{split}
\end{equation*}
After substituting the Christoffel  symbols (Eq.~(\ref{eq:cris})), $R^t_{rtr}$ is:
% \begin{equation}
% \begin{split}
% \implies& -\frac{1}{2}\partial_r(G^{tt}\partial_r G_{tt})-\frac{1}{2}G^{tr}G^{rt}(\partial_r G_{tt})^2+\frac{1}{2}G^{tt}G^{rr}(\partial_r G_{tt})(\partial_r G_{rr})\\
% &-\frac{1}{2}G^{tt}G^{rt}(\partial_r G_{tt})(\partial_r G_{tr})\\
% \implies & \neq 0,
% \end{split}
% \label{eq:re1}
% \end{equation}
\begin{align}
\implies\;&
-\frac{1}{2}\partial_r\!\left(G^{tt}\partial_r G_{tt}\right)
-\frac{1}{2}G^{tr}G^{rt}\left(\partial_r G_{tt}\right)^2 \nonumber\\
&\quad
+\frac{1}{2}G^{tt}G^{rr}
\left(\partial_r G_{tt}\right)\left(\partial_r G_{rr}\right)
\nonumber \\ &\quad-\frac{1}{2}G^{tt}G^{rt}
\left(\partial_r G_{tt}\right)\left(\partial_r G_{tr}\right)\neq 0
\label{eq:re1}
\end{align}
and $R^r_{ttr}$ is:
\begin{equation*}
\begin{split}
R^r_{ttr}=&-R^r_{trt}=-\partial_r\Gamma^r_{tt}+\Gamma^r_{t\lambda}\Gamma^{\lambda}_{rt}-\Gamma^r_{r\lambda}\Gamma^{\lambda}_{tt}\\
\implies& -\partial_r\Gamma^r_{tt}+\Gamma^r_{tr}\Gamma^r_{rt}+\Gamma^r_{tt}\Gamma^t_{rt}-\Gamma^r_{rr}\Gamma^r_{tt}-\Gamma^r_{rt}\Gamma^t_{tt}
\end{split}
\end{equation*}
Similarly after substituting Christoffel  symbols (Eq.~(\ref{eq:cris})), $R^r_{ttr}$ becomes:
% \begin{equation}
% \begin{split}
% \implies& \frac{1}{2}\partial_r(G^{rr}\partial_r G_{tt})-\frac{1}{2}G^{rr}G^{tt}(\partial_r G_{tt})^2+\frac{1}{2}G^{rr}G^{rt}(\partial_r G_{tt})(\partial_r G_{tr})\\
% &-\frac{1}{2}(G^{rr})^2(\partial_r G_{rr})(\partial_r G_{tt})\\
% \implies & \neq 0
% \end{split}
% \label{eq:re2}
% \end{equation}
\begin{align}
\implies\;&
\frac{1}{2}\partial_r\!\left(G^{rr}\partial_r G_{tt}\right)
-\frac{1}{2}G^{rr}G^{tt}\left(\partial_r G_{tt}\right)^2
 \nonumber\\ &\quad +\frac{1}{2}G^{rr}G^{rt}
\left(\partial_r G_{tt}\right)\left(\partial_r G_{tr}\right)
\nonumber\\
&\quad -\frac{1}{2}\left(G^{rr}\right)^2
\left(\partial_r G_{rr}\right)\left(\partial_r G_{tt}\right)\neq 0
\label{eq:re2}
\end{align}
All other components of the Riemann curvature tensor $[R^\rho_{\sigma \mu \nu}]$ except in Eq.~(\ref{eq:re1}), Eq.~(\ref{eq:re2}) are zero by anti-symmetries involving either $\rho=\sigma$ or $\mu=\nu$. The non-zero components (Eq.~(\ref{eq:re1}), Eq.~(\ref{eq:re2})) obtained thus implies a curved manifold represented by $G_{\mu \nu}$ which $\psi'$ sees for propagation. The kind of embedded acoustic spacetime within the fluid, unlike the outer static and symmetric spacetime (Eq.~(\ref{eq:1})) emerges as a result of the small deviations in flow variables from their stationary state values as evident from above. However it is worth mentioning that the emergent acoustic spacetime is independent of actual spacetime geometry of fluid motion, and the same is true even if the accretion process occurs in a Minkowski spacetime as well.
\section{Stability analysis: Standing and Traveling wave}
\label{sec:stability}
Let us consider a trial wave solution for the linear perturbation of mass accretion rate $\psi'$ given by:
\begin{equation}
\psi'(r,t)=A_\omega (r) e^{-i \omega t},
\end{equation}
satisfying  Eq.~(\ref{eq:alemb}) and expanding it further gives:
\begin{equation}
\begin{split}
&\omega^2 A_\omega(r)f^{tt}+i\omega[\partial_r(A_\omega(r)f^{rt})+f^{tr}\partial_r A_\omega(r)]\\
&-\partial_r[f^{rr}\partial_r A_\omega(r)]=0.
\end{split}
\label{eq:46}
\end{equation}
As mentioned earlier here we deal with transonic Bondi solutions only for a stationary state in which the flow variables passes through a critical point where $u(r_c)=a_s(r_c)$ and thus gradually for $r<r_c$ the flow becomes supersonic. Firstly we will be dealing with standing waves corresponding to the perturbation $\psi'$ which must satisfy the conditions of zero disturbance $A_\omega(r)=0$ at the two boundaries or end points. We further completely constrain ourselves to flows in subsonic region only keeping in mind the continuity of the solution $\psi'$ corresponding to standing waves.
Hence, standing waves corresponding to $\psi'$ between $r_1$ and $r_2$ must satisfies the boundary condition $A_\omega(r_1)=A_\omega(r_2)=0$. Multiplying Eq.~(\ref{eq:46}) with $A_\omega$ and integrating gives us:
\begin{equation}
\omega^2=-\frac{\int^{r_2}_{r_1} f^{rr}(\partial_r A_\omega)^2 dr}{\int^{r_2}_{r_1} f^{tt}A^2_\omega dr}
\label{eq:omega}
\end{equation}
It is clear from above that the sign of $\omega^2$ depends on analytical form of $f^{rr}$ and $f^{tt}$ only and from Eq.~(\ref{eq:matrix}):
\begin{equation*}
f^{tt}=\Big\{-g^{tt}+(1-\frac{1}{c^2_{\rm so}})(v_{\rm o}^t)^2\Big\}\frac{g_{rr}v_{\rm o}c^2_{\rm so}}{v_{\rm o}^t}
\end{equation*}
\begin{equation*}
f^{rr}=\Big\{g^{rr}+(1-\frac{1}{c^2_{\rm so}})v^2_{\rm o}\Big\}\frac{g_{rr}v_{\rm o}c^2_{\rm so}}{v_{\rm o}^t}
\end{equation*}
For subsonic flows $u_{\rm o}^2(r) < c^2_{\rm so}(r) \implies \frac{g_{rr}v^2_{\rm o}}{1+g_{rr}v^2_{\rm o}} < c^2_{\rm so} \implies (1-\frac{1}{c^2_{\rm so}})v^2_{\rm o} > g^{rr}$. Hence $f^{rr} > 0$ for subsonic flows only and similarly putting the expression of $v_o^t$ (Eq.~(\ref{eq:lor})) in $f^{tt}$ gives $-\{g^{tt}+g^{tt}\frac{1-c^2_{\rm so}}{(1-u_{\rm o}^2)c^2_{\rm so}}\} < 0$ . Use of these conditions implies that $\omega^2>0$ (Eq.~(\ref{eq:omega})) for subsonic flows with two real opposite roots thereby ensuring the stability of the standing waves corresponding to $\psi'$.
We now analyze the traveling waves corresponding to $\psi'$ which are characterized by high frequency waves propagating to large distances. We approximate the amplitude of $\psi'$ with a WKB like series approximation by:
\begin{equation}
A_{\omega}(r)=exp\Big\{\sum^\infty_{n=-1}\frac{k_n(r)}{\omega^n}\Big\}
\end{equation}
Putting $A_{\omega}$ in Eq.~(\ref{eq:46}) and equating the coefficients of $\omega,\omega^2$ further determines $k_0$. Further the leading coefficient $e^{k_0(r)}$ after some simplifications yields the amplitude of the perturbation $|\psi'|$ given by:
\begin{equation}
\begin{split}
&|\psi'| \sim e^{k_0(r)}\\
&\implies \Big(\frac{1+g_{rr}v^2_{\rm o}}{g_{rr}v^2_{\rm o}c^2_{\rm so}}\Big)^{\frac{1}{4}}
\end{split}
\label{eq:49}
\end{equation}
where $k_0(r)=-\frac{1}{2} \ln{\left(\sqrt{(f^{tr})^2-f^{rr}f^{tt}}\right)}$.
Eq.~(\ref{eq:49}) clearly shows non-divergent amplitude  $|\psi'|$ since for a spherical flow $v_o \neq 0$ except at infinity and $c_{\rm so} \neq 0$, thereby establishing the stability of the traveling waves in static and spherically symmetric spacetime.
For Schwarzschild spacetime : $g_{tt}=g_{rr}^{-1}=\Big(1-\frac{2}{r}\Big)$ (Eq.~(\ref{eq:1})) and putting it in above yields:
\begin{equation}
|\psi'|=\Big(\frac{1-\frac{2}{r}+v^2_{\rm o}}{v^2_{\rm o} c^2_{\rm so}}\Big)^{\frac{1}{4}}
\end{equation}
where EoS of the fluid determines the analytical expression of sound speed $c_{\rm s}^2$ (Eq.~(\ref{eq:speed})).

We further take the radial derivative of the above equation to get an insight of the varying amplitude of the perturbation with radial distance $r$:
\begin{equation}
\partial_r |\psi'|\sim \frac{2}{v^2_{\rm o} c^2_{\rm so} r^2}-\frac{4(1-\frac{1}{r})}{c^3_{\rm so}}\partial_r c_{\rm s}-\frac{2-\frac{4}{r}}{v^3_{\rm o}}\partial_r v_o
\label{eq:amp}
\end{equation}
We consider that both sound speed and fluid velocity should increase monotonically  (Fig. \ref{fig_mach}) towards origin for the transonic accretion solution implying both  $\partial_r v_o, \partial_r c_{\rm s} < 0$. Hence $\partial_r |\psi'(r)| > 0$ (Eq.~(\ref{eq:amp})), signifying monotonically shrinking behaviour of perturbation amplitude with decreasing radial distance towards origin. It is worthwhile to mention that Eq.~(\ref{eq:amp}) suggests the same effect of perturbation growth radially for spherically symmetric accretion in Minkowski spacetime, where, $g_{tt}=g_{rr}=1$ (Eq.~(\ref{eq:1})).
\section{Conclusions}
\label{sec:conc}
In this study, we investigated the dynamics of spherically symmetric, general relativistic accretion onto a Schwarzschild black hole, specifically considering a multi-component fluid with a radially varying adiabatic index. By constructing stationary, integral transonic solutions, we mapped the flow’s behavior through phase portraits in the Mach number–radial distance plane. Our dynamical systems analysis confirms that the physically relevant transonic critical points are of the saddle type, a result consistent with the expected behavior of inviscid flows in these environments.

The linear perturbations in the stationary transonic Michel flow with a varying adiabatic index behave exactly like a massless scalar field propagating through curved spacetime. This mathematical correspondence allowed us to identify an effective acoustic metric embedded within the flow, where the acoustic horizon aligns perfectly with the sonic surface. We visualized the causal structure of this emergent spacetime using Carter–Penrose diagrams and calculated the acoustic surface gravity as a function of the flow's conserved specific energy. Furthermore, the presence of non-vanishing Riemann curvature components demonstrates that this acoustic spacetime is genuinely curved. Our analysis also shows that these stationary solutions remain robustly stable when subjected to both standing and traveling linear perturbations.

It is to be noted that every analogue model found in the literature has
usually been obtained by linearly perturbing transonic (classical or quantum)
fluids.
This may suggest that the analogue gravity phenomenon is an artifact of the
linear perturbation.
That such an interpretation is not true has very recently been
proved~\cite{karan_das} by constructing the emergent spacetime through
the nonlinear perturbation of large-scale astrophysical flow under strong
gravity, which reinforces the belief that analogue gravity is a far more
deeply rooted phenomenon in nature than was previously thought.
In our next work, we will present the nonlinear perturbative analysis of
the Michel flow with variable $\Gamma$.
\section*{Acknowledgements}
SG would like to thank HRI for funding a visit to HRI, where a part of this work was completed.

\end{document}